\documentclass{article}
\usepackage{graphicx}
\usepackage{subfig}
\usepackage{setspace}
\usepackage{physics}
\usepackage{float}
\usepackage{authblk}
\usepackage[dvipsnames]{xcolor}
\usepackage{ragged2e}
\begin{document}

\title{Epsilon-Near-Zero Grids for On-chip Quantum Networks}

\author[1,*]{Larissa Vertchenko }
\author[1]{Nika Akopian}
\author[1]{Andrei V. Lavrinenko }
\affil[1]{Department of Photonics Engineering, Technical University of Denmark, {\O}rsteds Plads 345V, DK-2800 Kongens Lyngby, Denmark}

\affil[*]{lariv@fotonik.dtu.dk}

\flushbottom
\maketitle

\begin{abstract}
Realization of an on-chip quantum network is a major goal in the field of integrated quantum photonics. A typical network scalable on-chip demands optical integration of single photon sources, optical circuitry and detectors for routing and processing of quantum information. Current solutions either notoriously experience considerable decoherence or suffer from extended footprint dimensions limiting their on-chip scaling. Here we propose and numerically demonstrate a robust on-chip network based on an epsilon-near-zero (ENZ) material, whose dielectric function has the real part close to zero. We show that ENZ materials strongly protect quantum information against decoherence and losses during its propagation in the dense network. As an example, we model a feasible implementation of an ENZ network and demonstrate that information can be reliably sent across a titanium nitride grid with a coherence length of 434 nm, operating at room temperature, which is more than 40 times larger than state-of-the-art plasmonic analogs.Our results facilitate practical realization of large multi-node quantum photonic networks and circuits on-a-chip.


\end{abstract}

\begin{figure}[H]
\centering
	\includegraphics[scale=0.3]{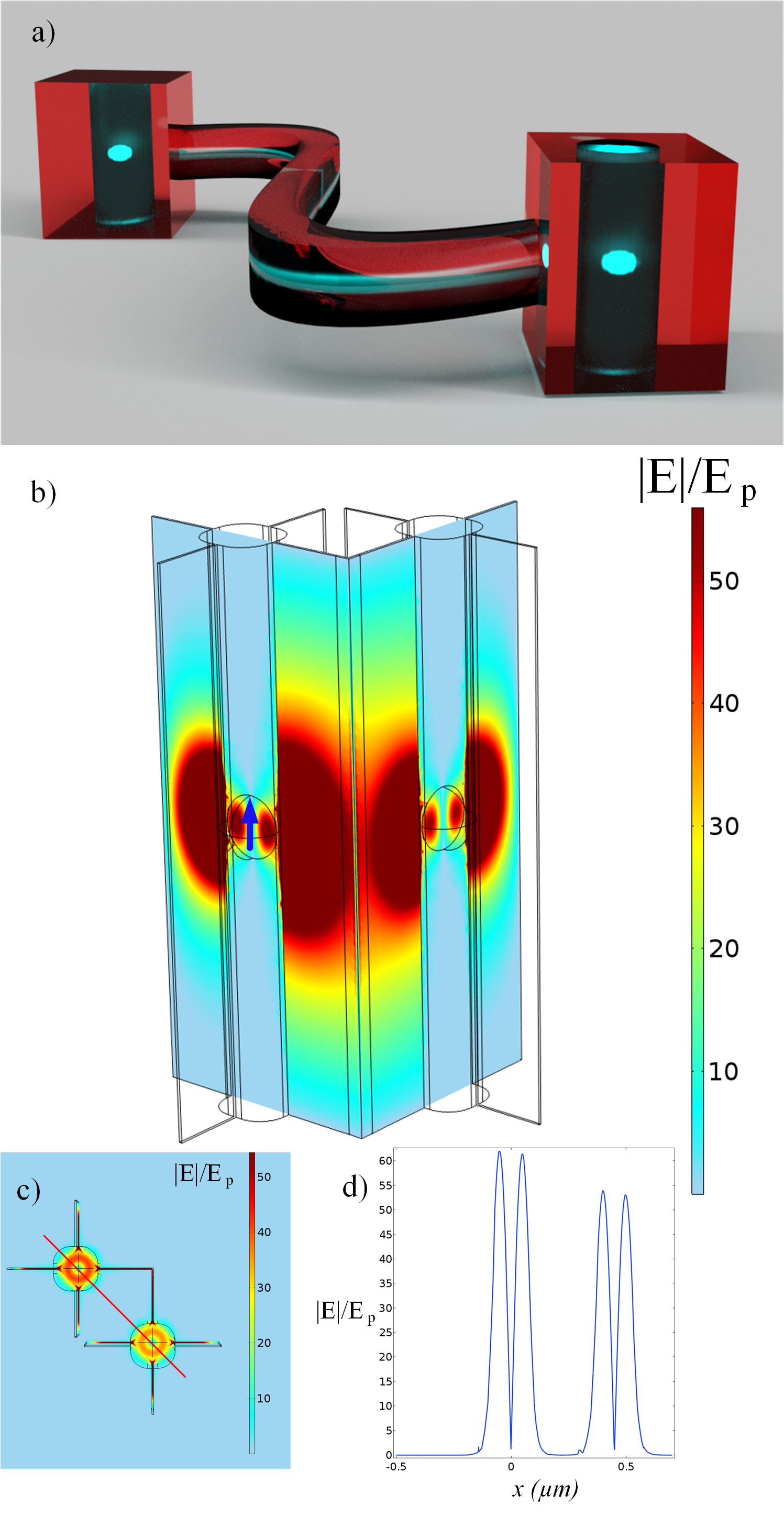}
	\caption{\textbf{Design concept of an ENZ quantum network and 3D simulations of the bent waveguide filled with ENZ material}.(a) Illustration of a QE (cyan) enclosed by a dielectric cylinder (dark blue), embedded in an ENZ waveguide with variable cross section (red), showing the possibility of interaction between distant QEs through the supercoupling effect. (b) The structure is composed of spherical air cavities, of radius 110 nm, embedded in the ENZ material, of $2 \mu m$ height and with a quantum emitter at the center (blue arrow). The normalized electric field profile is coded by the color scheme. For better visualization the gold layer is not displayed.c) Top view with a cut line linking the two cavities(red), (d) Electric field profile along the cut line.
\label{Fig1}}
\end{figure}

\thispagestyle{empty}
{\setstretch{2.0}
As described by H.J. Kimble \cite{1}, a Quantum Network (QN) is composed of three main elements: the nodes where the quantum information is generated, the channels which transport quantum states across the system and distribute entanglement between nodes and last, the light-matter interface for entanglement generation. The ultimate goal of on-chip photonic quantum technology will be met with the building of a network, where for instance,  entanglement can be coherently distributed \cite{2}. Current on-chip solutions involve the dielectric circuitry where dense integration is challenging. The alternative plasmonic networks can be arranged with a high density but they suffer from short coherence lengths on the scale of 1-10 nm \cite{3}.



Here, we exploit the epsilon-near-zero (ENZ) materials to overcome both limitations. Propagation of electromagnetic waves in ENZ materials exhibiting a close-to-zero relative permittivity has been an intense subject of research through recent years\cite{4}. One of their interesting feature is that waves are able to propagate in a subwavelength waveguide with acute bends almost without reflection losses. Such phenomenon is known as the supercoupling effect or tunneling \cite{5,6}. As the wavelength in an ENZ material is extremely long, the phase of the eigenmode is almost constant, allowing, for example, wavefront shaping \cite{7} for imaging applications. Another interesting feature of the ENZ materials is pronounced enhancement of nonlinearities \cite{8, 9}. It was reported that ENZ can also facilitate control over emission and interaction of quantum emitters (QE)\cite{10} embedded in an ENZ cavity, and that emitted photons could hold substantial entanglement over large distances.


Recently a quantum photonic platform capable of generation of multidimensional (16x16) entanglement has been experimentally demonstrated on a silicon chip \cite{11}. However, the key elements of this circuit have footprints in sizes of few micrometers, or even dozens. So making a denser grid is a real challenge for conventional photonic materials like silicon. It is well known that subwavelength sizes, abrupt changes of waveguide cross section, and presence of acute bends provoke back scattering and radiation of the mode, degrading its quality and affecting coherency. To push for the much smaller footprints of the circuit elements we propose to configure a QN with an ENZ material. Such an ENZ-based QN harvests on the supercoupling effect for synchronous excitation of multiple distributed QEs (see Supplementary Section 'Supercoupling theory'). In the ideal case when the ENZ is lossless (the imaginary part of the relative permittivity is equal to zero) the  mode at the output port is exactly the same as the input port having the footprint reduced in hundreds times in comparison with the Si-based elements \cite{11}.

Currently quantum dots (QDs) are considered as single photon emitters, which can be used as a source of coherently created photon pairs \cite{12}. Quantum dots can be naturally implemented in different epitaxially grown structures like nanowires and micropillars, which provide highly efficient channels of light outcoupling. These single photons can transport quantum information, encoded either in polarization or time-bin qubits. Since our network design supports the propagation of only TM modes \cite{7}, we can not use polarization qubits. We, therefore, propose to encode quantum information in time-bin qubit \cite{13} described by the superposition of two pulses, $ \ket{\psi}= \alpha\ket{early}+\beta e^{i\phi}\ket{late}$, where $\alpha$ and $\beta$ are general probability amplitudes and $\ket{early}$ and $\ket{late}$ represent the state of the pulses separated in time \cite{14}. Inherently, the immunity of the time-bin qubit during propagation directly depends on property of keeping the phase difference $\phi$  between these two pulses constant.

The principle of operation of a dense grid of ENZ channels is demonstrated with a two-cavities example in Fig.\ref{Fig1}a. To allow photon emission a quantum emitter has to be placed in dielectric insertion \cite{15}. The high contrast between the refractive indices of the ENZ and dielectric effectively forms a cavity. The QE emission can be enhanced by fitting the size of the cavity to the resonance conditions \cite{16}. Aiming to match the emission spectrum of a typical GaAs quantum dot \cite{17} we chose a wavelength of $780\ nm$. Then, the radius of the spherical cavity, $r = 110\ nm$, was optimized to achieve a magnetic dipolar resonance (see Supplementary Section `Optimization'). Knowing that QDs can have the size of just a few nanometers, the structure is considered feasible for nanofabrication. The channel width, length and height are flexible parameters, and we fix them to $10\ nm$, $1\ \mu m$ and $2\ \mu m$, respectively.  Outside the waveguide a $100\ nm$ thick layer of gold was used to prevent leakage in the environment\cite{6}. We present results for the full 3D simulations in Fig. \ref{Fig1}b). The ENZ material for illustration purposes was chosen to have a very small permittivity for both real and imaginary parts ($\varepsilon = 10^{-3} + i10^{-3} $). The normalized electric field profile along the straight line connecting both cavities displays a reduction in the peak value of the amplitude by approximately $14 \% $, as shown in Fig\ref{Fig1}c) and Fig\ref{Fig1}d). We point out this reduction is subjected to sizes and configuration of the channels.
 
 The electric field produced by the QE simulated as a point source, with dipole moment $\boldsymbol{d}$, placed in the center of the left cavity in Fig.\ref{Fig1}b is transmitted with high efficiency through a deeply subwavelength bent ENZ waveguide with negligible losses, which makes it possible to excite another emitter with the same emission frequency in the second cavity. To this end, we compute the decay rate of two emitters ($\boldsymbol{d_1},\boldsymbol{d_2}$) due to coupling, $\Gamma_{21}= 2k_0^2/ (\hbar\varepsilon_0) \boldsymbol{d_2} \cdot Im(\boldsymbol{G(r_2,r_1, \omega)}) \cdot \boldsymbol{d_1} $, where $(\boldsymbol{G(r_2,r_1, \omega)})$ is the Green's electric field tensor\cite{16}, $k_0$ is the wave number and $\varepsilon_0$ the free space permittivity. The frequency shift due to dipole-dipole interactions (Lamb shift) is calculated according to $\Delta \omega_{21}= -k_0^2/ (\hbar\varepsilon_0) \boldsymbol{d_2} \cdot Re(\boldsymbol{G(r_2,r_1, \omega)}) \cdot \boldsymbol{d_1} $.The plot for such cooperative behaviour is depicted in Fig.\ref{Figcoupling}, where the decay rate and Lamb shift related to coupling are normalized by the free space decay rate (See Supplementary Section 'Dispersion Model').


The transport of quantum states across the networks \cite{1} suffers from decoherence as a result of the interaction with the environment. Therefore, one of the major current challenge with QNs is to attain coherent transfer of quantum states from spatially-separated quantum emitters \cite{18}. A small wave vector supported by the ENZ materials \cite{6} helps to have a constant phase difference between the wavefronts of the signals \cite{19}. The fact that all conducting electrons of the ENZ material oscillate synchronously, leads to coherent processes of quantum emitters communication on different distances and eventually  supports coherent control over light-matter interactions\cite{1}.   

To evaluate the reasonable dimensions of a QN the coherence length should be assessed. This length is connected with the coherence time, which determines the interval when the phase difference between the signals stays constant. To calculate the coherence time of the system one should find the relaxation time of the collective electron oscillations \cite{20}, which is related to the imaginary part of the permittivity by the full width half maximum of the loss function (see Supplementary Section `Temporal coherence'). Even with small losses the temporal part of the electric field is exponentially damped, which, in turn, affects the coherence time. 

For realistic analysis we use the dispersion curve of a silicon carbide (SiC), which achieves the ENZ regime with permittivity $\varepsilon = 0 + 0.1i$ at the wavelength of 10.3 $\mu m$ \cite{16}. Using the equations for the autocorrelation function and the degree of temporal coherence \cite{21}, we found the coherence time of $1.061\times 10^{-12} s$. Considering that the mode propagates with a phase velocity equals to $\omega/k$ gives us the coherence length  of $1.4\ mm$. As a alternative to SiC on visible frequencies we challenge titanium nitride (TiN) with the ENZ point at the wavelength of $667\ nm$ with permittivity $\varepsilon = 0 + 4i$ \cite{22}(see Supplementary Section `SiC and TiN permittivity curves').
Then the coherence time is $ 2.08\times 10^{-15} s$, providing the coherence length of $434\ nm$. While it is rather short, comparing it with the coherence length in noble metals, which is typically in order of $1-10 \ nm$ \cite{3}, it exposes a considerable improvement of at least 40 times. Such values imply that the time-bin qubit $\ket{\Phi}$ generated by the QE would be able to propagate a long distance before collapsing into the early or late states, giving enough room to implement logic operations inside the network \cite{23}, as well as, opening the possibility for multipath entanglement \cite{24,25}.


The results for the bent waveguide motivated for expansion of the system with multiple crossing channels, forming what we actually call an ENZ network. Computation-wise we reduce our analysis to the two-dimensional case (2D), which is still able to exhibit most essential features of the network. In Fig \ref{Fig2} we show results for the ENZ grid consisting of 5 x 5 identical cavities occupying circa a 15 x 15 $\mu m^2$ area. To work in the optimal conditions the radius of the cavities was chosen to be $310\ nm$ . The point source is located in the central cavity (blue arrow in Fig.\ref{Fig2}). The field intensity distribution shown in Fig.\ref{Fig2}(a) visually confirms the equal expansion of fields in all cavities even not directly connected to the central one (See Supplementary Section `Intensity profile'). The phase is preserved within the whole network, as depicted in Fig.\ref{Fig2}(b), and is in the range of the coherence length of a material close in its optical properties to the ENZ point of SiC. In Fig.\ref{Fig3} we illustrate the possibility of a dense grid within the coherence length.  By using a radius corresponding to the coherence length of 1.4 mm we could estimate a maximum number of nodes that would fit inside the low loss grid. For an unitary cell with 2.089 $\mu m$ of length we found a value of approximately $1.41\cross10^6$ nodes. While state-of-art single crossings have dimensions of around 30 $\mu m$ \cite{26}, we were able to decrease this value by 15 times, which represents a breakthrough in terms of scalability.


The dense ENZ grid of cavities can be easily extended further. For example,  for a square ENZ grid of 15 x 15 cavities (see Supplementary Section `Bigger Networks') the electric field decays much slower than in the same size network of cavities but solely filled with a material of $\epsilon = 1$, such as air, (see Supplementary Section 'Curve Fitting'), see comparison in Fig.\ref{Fig4}. There is still a considerable signal in the furthest cavity of the ENZ network, whereas the field in the air-filled network is four orders of magnitude less.
 
 
Therefore, a single QE can access all other distant emitters in the whole grid realizing the favorite scenario for multidimensional entanglement. To illustrate this we embedded gold cylinders of radius $70\ nm$ in each cavity  (Fig.\ref{Fig5}). The particles are placed in the sites with the highest electric field and the active QE is positioned in the central cavity. After some time all gold cylinders exhibit an intensity distribution characteristic for a dipole resonance, oscillating in phase, confirming the possibility of simultaneous excitation of numerous distant particles connected through the dense ENZ grid. This feature is well suited for the QN, because the equal phase electric field delivery in each of the cavity can help to acquire collective entanglement of photons emitted by an array of quantum emitters.

One specific limitation of the ENZ network is that it demands low losses, since intrinsic losses are responsible for significant deterioration of the signal and have the greatest influence on the coherence properties of ENZ \cite{27}. Several alternatives have been proposed in order to mitigate the problem of losses, such as, usage of all-dielectric metamaterials \cite{28}, operating photonic crystals at Dirac's triple point \cite{29}, loss compensation by gain material, i. g., fluorescent dyes \cite{30,31} or cooling waveguides to cryogenic temperatures. Further analysis of their suitability in QNs is required. 

In conclusion, we introduced the concept of ENZ grid for on-chip QNs, where we exploited the supercoupling effect on systems of QEs. Strong coupling between distant emitters and high confinement inside bent channels present a great potential for the design of shape-flexible on-chip QNs with the density of elements in hundreds of times exceeding these available with Si photonics. Moreover, due to the long coherence length, the dense ENZ grids acquire clear bonus against networks from conventional plasmonic materials. We found the coherence length of TiN waveguides of $434$ nm for the wavelength of $667$ nm, which is close to typical operational wavelengths of quantum dots. SiC exhibits even higher lengths, about 1.4 mm, however, at the wavelength of $10.3$ $\mu$m. The fast progress in utilization of the mid-IR range gives certain promises for QNs extension to this domain too. Besides, the homogeneously distributed excitation of nanoantennas in classical grid systems can be exploited in sensing applications \cite{32}, and here the $10.3$ $\mu$m networks can be heavily employed. Our findings can unprecedentedly facilitate the fields of quantum photonics and propose a feasible implementation in a short-term perspective. 

}
\bibliography{bibliography}
\section*{Methods}

The system was modeled by the finite element method, using the commercially available software COMSOL \cite{33}.

\section*{Acknowledgements}

The authors thank N. Engheta for discussions and E. Shkondin for providing TiN characterization data.

\section*{Author contributions statement}
All authors conceived the problem. L.V. designed the structures, performed simulations and data analysis. All authors discussed the results and wrote the manuscript. 

\section*{Competing Interests}

The authors declare no competing interests.

\newpage

\begin{figure}[H]
\centering
	\includegraphics[scale=0.4]{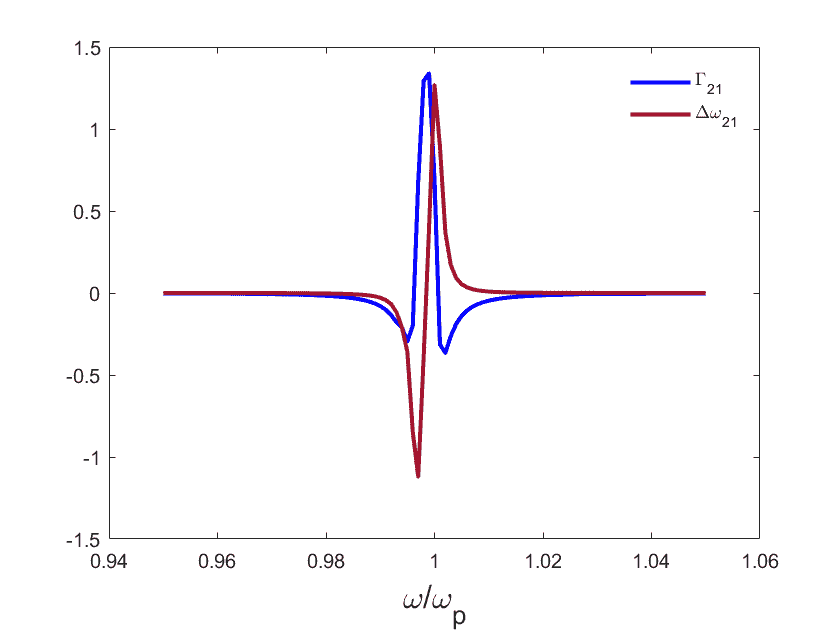}
	\caption{\textbf{Coupling of two emitters}.Decay rate due to coupling of two dipoles normalized by the free space decay rate, as a function of the frequency normalized by the plasma frequency $\omega_p$. The dipoles were located in cavities with a 2 $\mu m$ distance from each other.
\label{Figcoupling}}
\end{figure}

\begin{figure}[H]
\centering
	\includegraphics[scale=0.15]{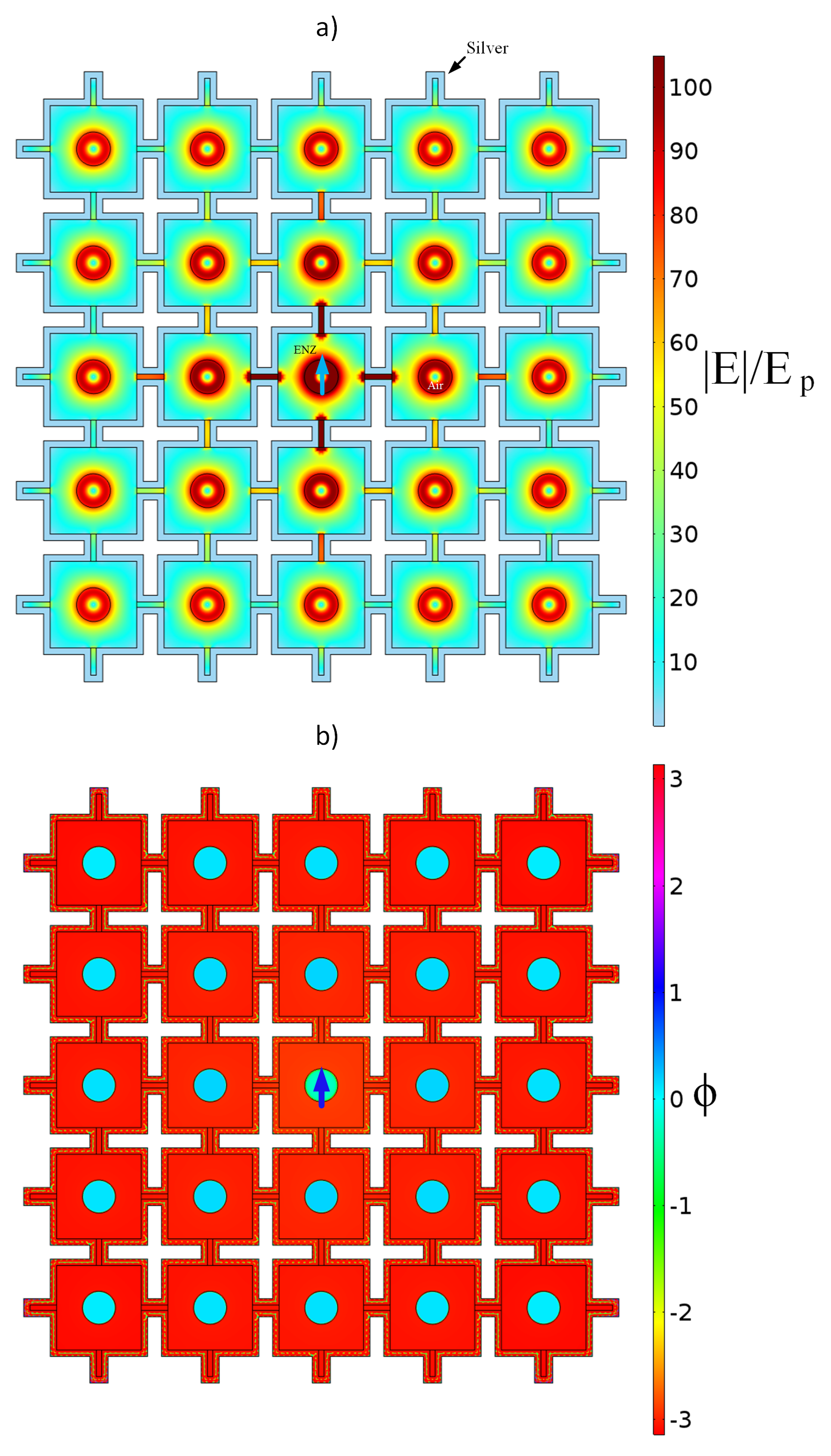}\advance\rightskip-2cm \centering
	\caption{\textbf{2D ENZ Quantum Network}. 2D simulations of an electric field emitted by a point source, placed inside the middle cavity, and phase distribution in a ENZ network. (a) Normalized electric field of a quantum emitter placed at the center of the network, represented by the blue arrow.(b) Phase distribution of the magnetic field $H_z$. \label{Fig2}}
\end{figure}

\begin{figure}[H]
	\includegraphics[scale=0.25]{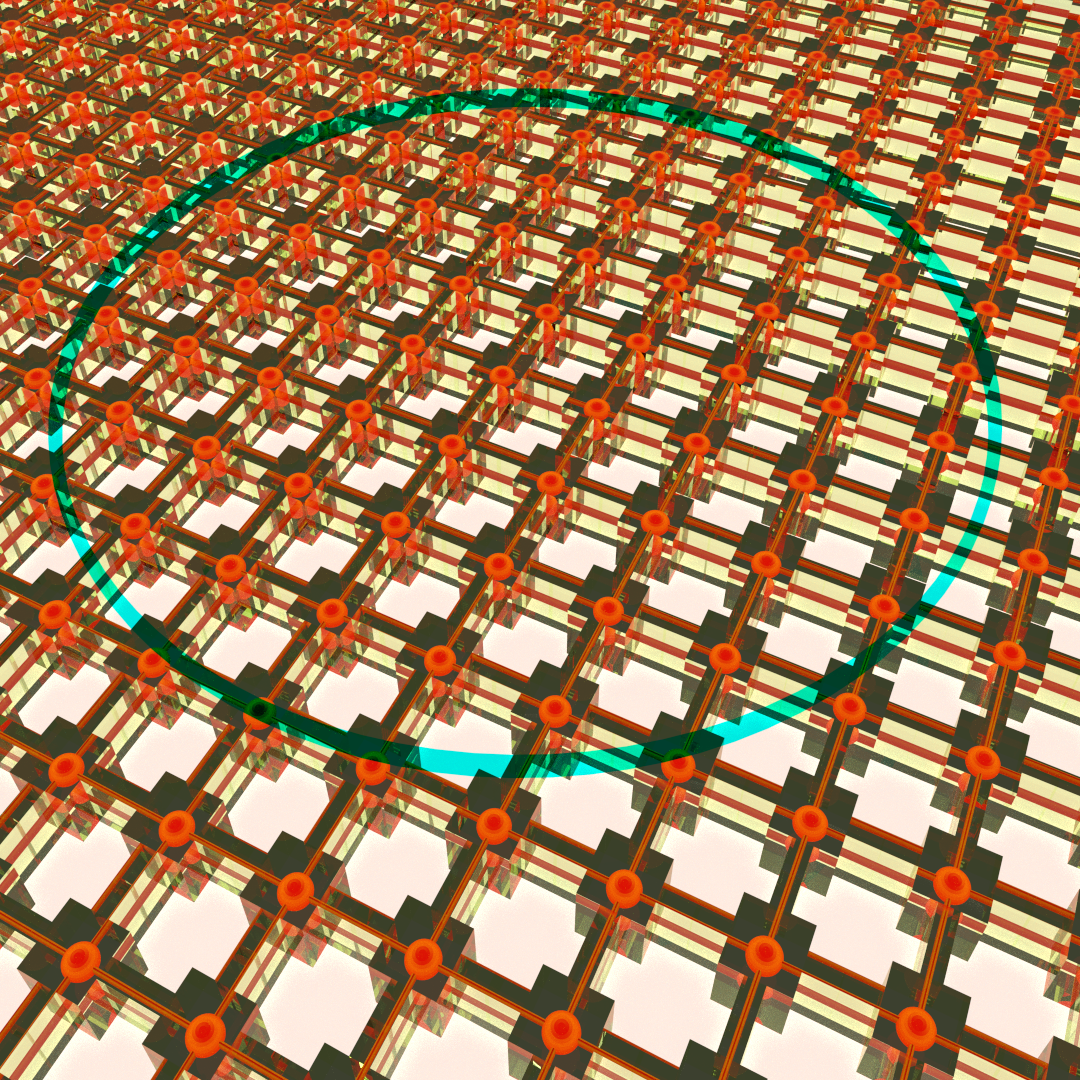}
    \centering
	\caption{\textbf{Illustration of a dense ENZ  quantum network}.The electromagnetic field is artistically depicted by the red color. The blue circle represents the radius of coherence for a quantum emitter placed on its center. The coherence and entanglement properties can only be preserved within the circle. Although the signal can propagate further this limit, quantum information would be lost due to decoherence. 
    \label{Fig3}}
\end{figure} 

\begin{figure}[H]
\centering
	\includegraphics[scale=0.3]{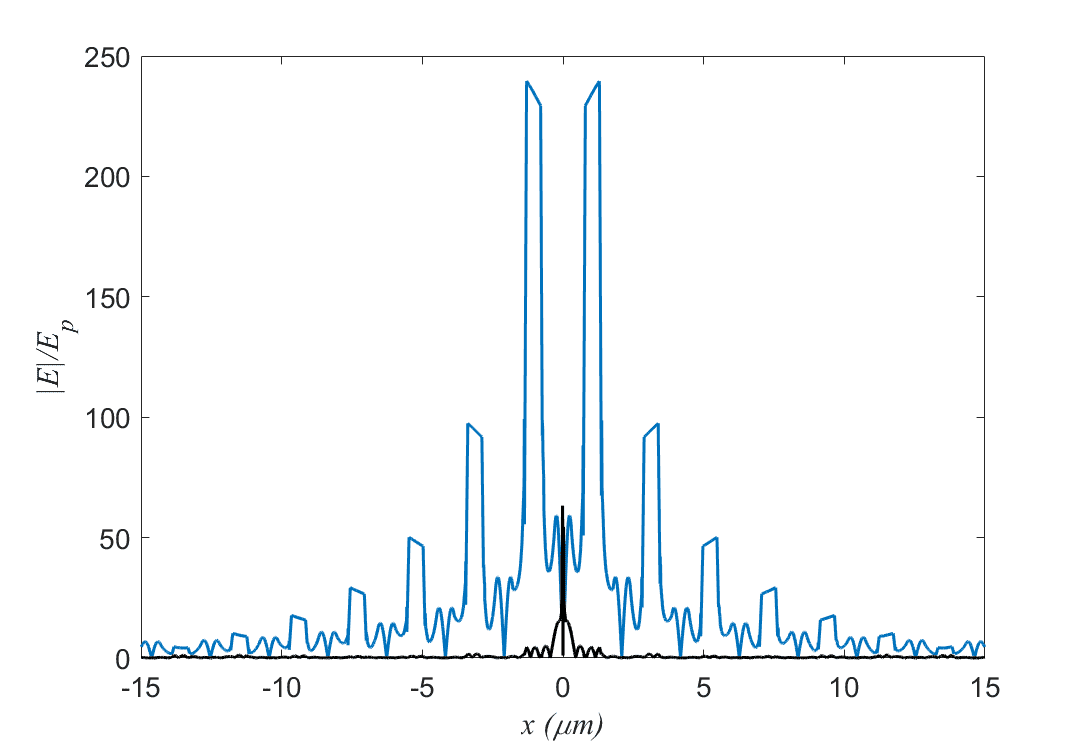}\par
	\caption{\textbf{Comparison between air and ENZ waveguides}.(a) Normalized (to the amplitude of the point source) electric field, as a function of the distance from the middle cavity, for an ENZ network (blue) and air network (black), both with the same number of cavities.
    \label{Fig4}}
\end{figure}

\begin{figure}[H]
	\includegraphics[scale=0.4]{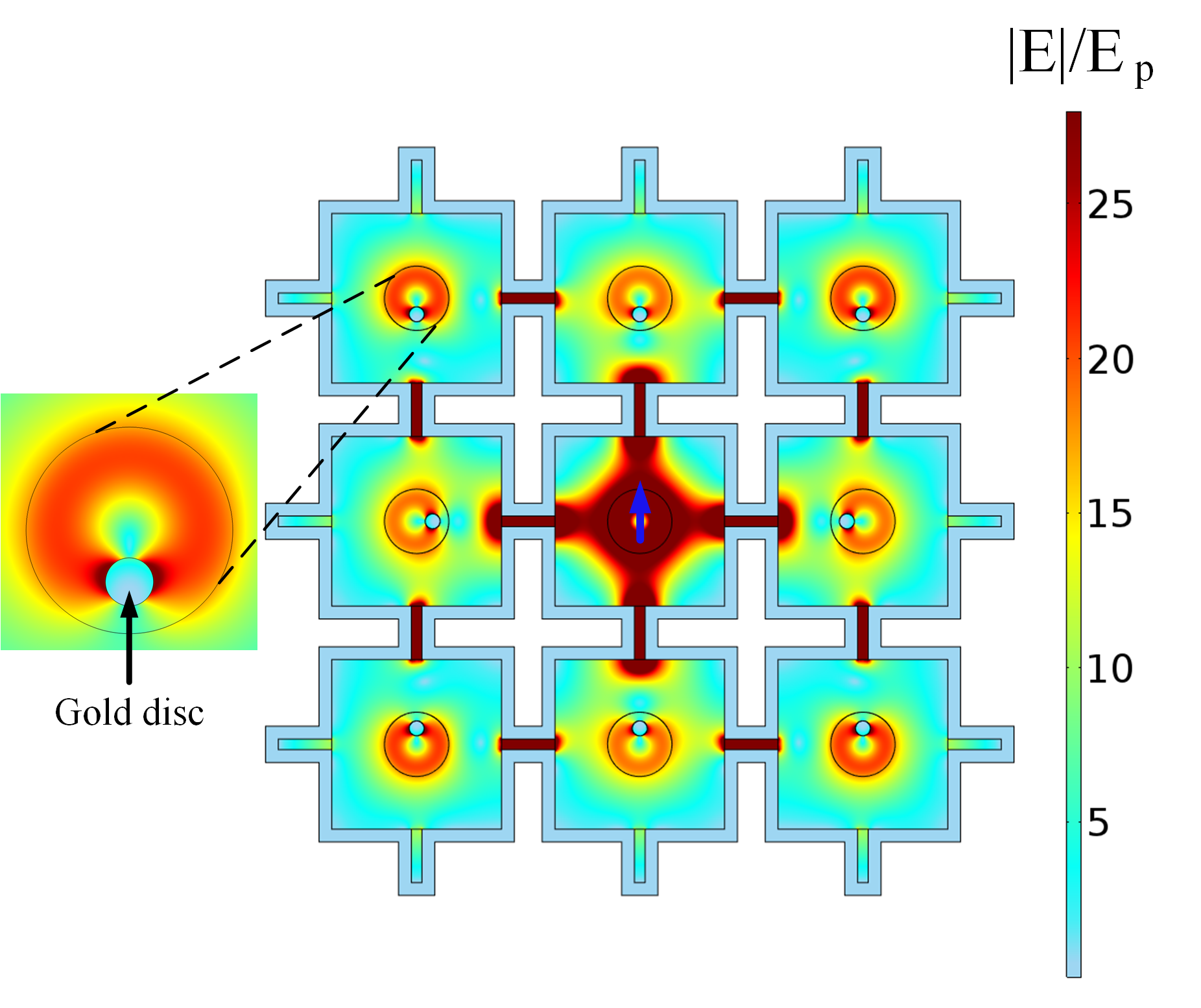}
    \centering
	\caption{\textbf{Excitation of gold nanodiscs inside the ENZ network}.Normalized electric field distribution of an ENZ network with gold discs placed inside each cavity, showing a dipolar excitation response to the QE radiation. The blue arrow represents the QE position.
    \label{Fig5}}
\end{figure} 

\bibliographystyle{unsrt}

\end{document}